\documentclass[epjST]{svjour}
\usepackage{graphics}
\usepackage{amsmath}
\usepackage{amssymb} 
\begin{document}
\title{Exploring noise-induced chaos and complexity in a red blood cell system}
\author{Bo Yan\inst{1} \and Sayan Mukherjee\inst{2} \fnmsep\thanks{\email{msayan80@gmail.com}} \and Asit Saha\inst{3}}
\institute{Department of Information Engineering, Shaoyang University, Shaoyang 422000, China \and Department of Mathematics, Sikkim Manipal Institute of Technology, Sikkim Manipal University, Majitar, Rangpo, East-Sikkim 737136, India \and Department of Mathematics, Sivanath Sastri College, West Bengal, Kolkata 700029, India}

\abstract{
We investigate dynamical changes and its corresponding phase space complexity in a stochastic red blood cell system. The system is obtained by incorporating power noise with the associated sinusoidal flow. Both chaotic and non-chaotic dynamics of sinusoidal flow in red blood cell are identified by $0-1$ test. Furthermore, dynamical complexity of the sinusoidal flow in the system is investigated by heterogeneous recurrence based entropy. The numerical simulation is performed to quantify the existence of chaotic dynamics and complexity for the sinusoidal blood flow.
} 
\maketitle
\section{Introduction}
\label{intro}
Red blood cell (RBC) is one of the primary ingredients of human blood. The RBCs consist of hemoglobin covered by a very thin elastic membrane (cytoskeleton) \cite{REF1}. The cytoskeleton contains cytoplasmic fluid which controls the deformation of the cell and influence its microscopic flow \cite{REF2}. Under the normal condition, cytoskeleton deforms the RBC at a constant area and exhibits both elastic and viscous behaviour \cite{REF2}. However, perturbation in the membrane deformation can reveal complex fluid structure in the blood flow. In \cite{REF3}, a red blood cell system was proposed to study the oscillation in RBC. Further, its complex behaviour of RBC under sinusoidal flow was investigated in \cite{REF4} by applying the theory of nonlinear dynamics. \par
Nonlinear dynamics is a ubiquitous property of a time-dependent system \cite{REF5,REF6,REF7,REF8,REF9}. It can predict long-term behaviour of the systems \cite{REF6,REF7,REF8,REF9}. Various long-term dynamics can be observed in a system \cite{REF6,REF7,REF8,REF9,REF10}. Among these, chaotic dynamics is one of the most complex behaviour in the systems \cite{REF6,REF7,REF8,REF9}. In a chaotic state, system losses the information of the previous memory and becomes unpredictable \cite{REF6,REF7,REF8}. Chaos can be observed in both deterministic and stochastic systems \cite{REF6,REF7,REF8,REF9,REF10,REF11,REF12,REF13,REF14,REF15}. In the deterministic case, Lyapunov exponent is applied to recognize chaos in the system \cite{REF6,REF7,REF8}. On the other hand, $0-1$ test \cite{REF16,REF17,REF18,REF19} is one of the efficient methods to characterize chaos in a stochastic system \cite{REF19}. In \cite{REF3}, chaos has been already established under the variation in shear rate of the sinusoidal flow. Indeed, effect of noise on the same system is not explored. As noise is an inherent feature that appears with different form in the biological phenomena, the noise-induced long-term analysis will be one of the best possible studies to approximate future of the stochastic RBC system. Also, variation in the dynamics under the changes in the small axis of the RBC is not examined. So, we investigated chaotic and non-chaotic dynamics under the variation in noise strength and the small axis of the RBC. \par
The complexity in a system can further be investigated to point out disorder in the corresponding phase space \cite{REF20,REF21,REF22,REF23,REF24,REF25,REF26,REF27}. The disorder can be quantified by entropy proposed by Shannon in \cite{REF22,REF23,REF24,REF25,REF26,REF27}. Based on Shannon entropy, several measures were proposed to quantify the dynamical complexity \cite{REF22,REF23,REF24,REF25,REF26,REF27,REF28}. In \cite{REF28}, an entropy based on recurrence plot (RP) was proposed that can be applied in any dimensional system. RP describes such recurrence states which are belonging to a thresold based neighbourhood \cite{REF28,REF29}. As RP delineates two recurrent states as same, whatever the difference in the regions are, transition of local dynamics in the system can not be described \cite{REF30,REF31,REF32,REF33,REF34}. A new entropy based on heterogeneous recurrence (HR) was proposed in \cite{REF30}, which can successfully measure the complexity of local transition in the asymptotic dynamics \cite{REF31,REF34}.\par
This article is organized as follows: In section \ref{sec:1}, a stochastic red blood cell system is described. The noise is taken as $\frac{1}{f^{\beta}}$-noise with $\beta=1$. The reason for incorporating $\frac{1}{f}$-noise can be state as follows:\\
i) $\frac{1}{f}$-noise (known as pink noise) is a correlated noise, i.e; it has long-range dependence \cite{REF35,REF36}. It affects the long-term dynamics of the system. \\
ii) $\frac{1}{f}$- fluctuation can be observe in many biological system-heart, brain and even music, etc \cite{REF35,REF36}. As dynamics of RBC depends on the sinusoidal blood flow and the flow is control by human heart and brain, the effect of $\frac{1}{f}$-noise is an obvious phenomenon in the RBC system.\par
Then, dynamical variation is investigated by $0-1$ test method. The asymptotic dynamics are described by phase space and max-map analysis. Section \ref{sec:2} discusses complexity of the phase spaces based on heterogeneous recurrence and its entropy measure-heterogeneous recurrence entropy (HRP) \cite{REF30,REF31,REF32,REF33,REF34}. Further, dynamical complexity was investigated on our proposed stochastic RBC system. Finally, a conclusion is given in section \ref{sec:3}. 
\section{Investigating noise-induced chaos in blood flow}
\label{sec:1}
\subsection{Stochastic blood flow system}
\label{sec:11}
An autonomous blood flow system, proposed by Dupire et. al. [4], is considered to study motion of a RBC under a moderate shear flow is given by
\begin{equation} \label{eq1}
\begin{split}
\frac{dx}{dt} & = \frac{\dot{\gamma}_a \sin(z)}{2}(A \cos(2x)-1)-B \frac{dy}{dt}, \\
\frac{dy}{dt} & = C \dot{\gamma}_a (D \sin(2y)-\sin(z) \cos(2x)), \\
\frac{dz}{dt} & = \nu,
\end{split}
\end{equation}
where $x,y$ represent respective cell inclination angle with respect to the flow direction and the angle between the instantaneous position of a membrane element and its initial position at rest. Here, $z$ is represented by $z=\nu t$ connected to the sinusoidal flows $\dot{\gamma}=\dot{\gamma}_a \sin(\nu t)$ ($\dot{\gamma}_a$ being the shear rate of the sinusoidal flow) . The quantities $A,~B,~C,~D$ are given by $A=\frac{a_1^2-a_2^2}{a_1^2+a_2^2}$, $B=\frac{2 a_1 a_2}{a_1^2+a_2^2}$, $C=\frac{-f_3}{(f_2-(\frac{\eta_i}{\eta_0}) (1+(\frac{\eta_m}{\eta_i} \Omega_V)) f_1}$, $D=\frac{f_1}{2 f_3} \frac{\mu_m}{\eta_0 \dot{\gamma}_a} \frac{\Omega}{V}$. Here, $a_1$ and $a_2$ are the axes of the cell cross section shown in Fig.1 of \cite{REF4}. The quantities $\Omega, V$ and $\eta_0$ represent RBC volume, membrane volume and external suspending fluid viscosity respectively. The $f_1, f_2, f_3$ are known as geometrical constants described in \cite{REF3}. In \cite{REF3}, $f_1, f_2, f_3$ are given by 
$f_1 = \left[\frac{a_2^2-a_1^2}{a_1 a_2}\right]^2,~ f_2= f_1 \left[1-\frac{2}{\gamma_0^{'}a_1 a_2 a_3 (a_1^2+a_2^2)}\right], ~f_3= -\frac{2 (a_1^2-a_2^2)}{a_1^2 a_2^2 a_3 \gamma_0^{'} (a_1^2+a_2^2)},$
where $\gamma_0^{'}$ is calculated by 
\begin{equation} \label{eq2}
\gamma_0^{'}=\int_0^\infty \frac{d \mu}{(a_1+\mu)^{\frac{3}{2}}(a_2+ \mu)^{\frac{3}{2}}(a_3+\mu)^{\frac{1}{2}}}.
\end{equation}
Then, corresponding stochastic blood flow is described by incorporating Power noise $\Phi=\frac{1}{f^{\beta}}$ with the systen (\ref{eq1}). The obtained system is given by  
\begin{equation} \label{eq3}
\begin{split}
\frac{dx}{dt} & = \frac{\dot{\gamma}_a \sin(z)}{2}(A \cos(2x)-1)-B \frac{dy}{dt}, \\
\frac{dy}{dt} & = C \dot{\gamma}_a (D \sin(2y)-\sin(z) \cos(2x)), \\
\frac{dz}{dt} & = \nu+K \Phi,
\end{split}
\end{equation}
where $K$ represents noise strength. \par
For the numerical computation, one can fix $a_1=a_3=4 \mu m$, $\frac{\Omega}{V}=7.482 \times 10^{-2}$, $\eta_0=34 \time 10^{-3} Pa. s$, $\eta_i=10^{-2} Pa.s$, $\eta_m=0.7 Pa.s$, $\mu_m=1.6 Pa$ as chosen in \cite{REF4}.  
\subsection{Investigation of chaotic dynamics of the blood flow system}
\label{sec:12}
Chaotic and non-chaotic dynamics of sinusoidal flow in red blood cell (\ref{eq3}) are investigated under both the variation of $K \in [0,0.1]$ and $a_2 \in [1,1.5]$ respectively. To investigate this, $0-1$ test analysis is employed. In this analysis, a single solution component, say $s(j),~j=1,2,...,N$ ($N$ being the length of the component) is chosen. Then, $\{s(j)\}$ is transformed to
\begin{equation}\label{eq4}
p^c(n)=\sum_{j=1}^{n}s(j)\cos(jc),\quad q^c(n)=\sum_{j=1}^{n}s(j)\sin(jc),
\end{equation}
where $c \in (0,\pi)$ and $n=1,2,..,N$.\par
\begin{figure}[h!]
\resizebox{1.0\columnwidth}{!}{%
  \includegraphics{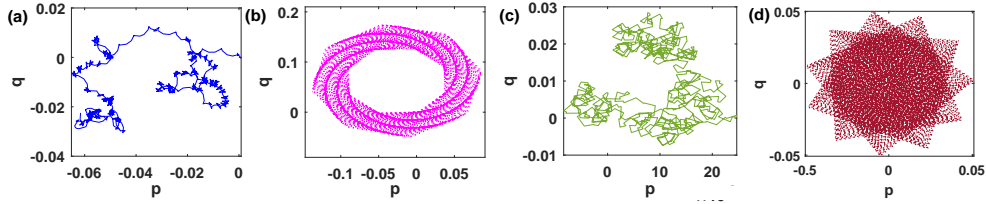} }
\caption{$pq$-plot analysis: (a)-(d) respectively $(p,q)$-clouds for the system (\ref{eq3}) with the respective $(K,a_2)=(0.01,1),~(0.01,1.5),~(0.06,1),~(0.045,1.5)$. In order to construct the $pq$-plots, we have considered $x$-component of the system (\ref{eq3}). The value of $c$ is taken by $[\frac{\pi}{5},\frac{4\pi}{5}]$.}
\label{fig:1}       
\end{figure}

A $2$D plot consisting $(p^c,q^c)$ points, known as $pq$-plot, can indicate both chaotic and non-chaotic dynamics \cite{REF16,REF17} of sinusoidal blood flow in the system (\ref{eq3}). In fact, Brownian motion like structure and regular geometry of the $pq$-plots identify chaotic and non-chaotic dynamics \cite{REF16,REF17,REF18,REF19} of sinusoidal blood flow, respectively. Fig.\ref{fig:1} shows some of the $pq$-plots for the system (\ref{eq3}). 
From Fig.\ref{fig:1}(a) and (c), it can be observed that the corresponding $pq$-plots contain Brownian motion like path. It indicates chaos in (\ref{eq3}) for $(K,a_2)=(0.01,1),~(0.06,1)$. On the other hand, regular geometric structures can be observed in both Fig.\ref{fig:1}(b) and (d) respectively. It implies existence of non-chaotic dynamics in (\ref{eq3}) with the respective $(K,a_2)=(0.01,1.5),~(0.045,1.5)$. To quantify dynamics of  sinusoidal blood flow, we implement diffusive and non-diffusive behavior of $p^c$ and $q^c$. Then a measure as been proposed in \cite{REF16}, called mean square displacement $M_c$ defined by
\begin{equation}\label{eq5}
M_c=\lim_{N\to\infty} \frac{1}{N}\sum_{j=1}^{N}[p^c(j+n)-p_c(j)]^2+[q^c(j+n)-q_c(j)]^2,
\end{equation}
where $n<<N$. \par
\begin{figure}[h!]
\resizebox{0.90\columnwidth}{!}{%
  \includegraphics{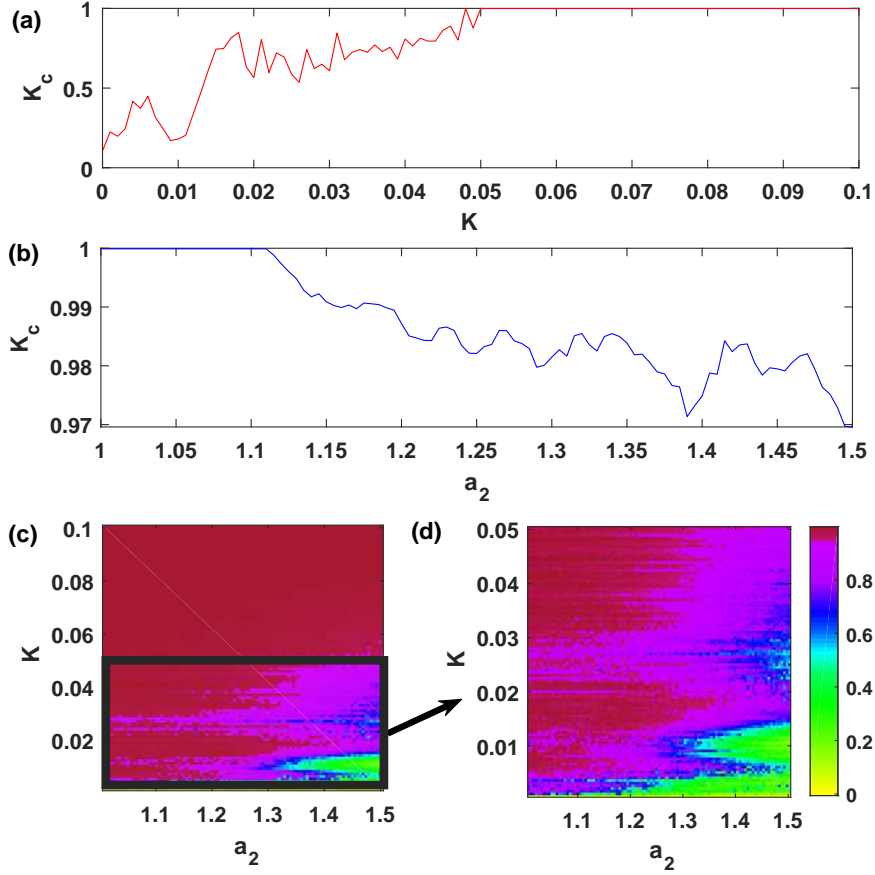} }
\caption{$0-1$ test analysis: (a) and (b) represents $K_c$ vs. $K \in [0,0.1]$ ($a_2=1.5$) and $a_2 \in [1,1.5]$ ($K=0.045$) for the system (\ref{eq3})respectively. (c) represents $[K_c(K,a_2)]$-matrix plot for the same system with $(K,a_2) \in [0,0.1] \times [1,1.5]$. The rectangular region, marked by black solid lines, zoomed in (d). The associated color bar indicates values of $K_c$. In each cases, values of $K_c$ are calculated from $x$-component of the system (\ref{eq3}) with length $N=15000$, $c \in [\frac{\pi}{5},\frac{4\pi}{5}]$.}
\label{fig:2}       
\end{figure}

  
The behavior of $M_c$ is calculated by the asymptotic growth of $M_c$ which is given by
\begin{equation} \label{eq6}
K_c=\lim_{n\to\infty} \frac{\log M_c(n)}{\log n}.
\end{equation}
The value of $K_c$ close to $1$ and $0$ indicates chaotic and regular dynamics respectively \cite{REF16,REF17}. Using (\ref{eq6}), we have investigated fluctuations of $K_c$ over the respective variation in $K \in [0,0.1]$ ($a_2=1.5$) and $a_2 \in [1, 1.5]$ ($K=0.045$). The corresponding oscillations are shown in Fig.\ref{fig:2}(a) and (b) respectively. 

\begin{figure}[h!]
\resizebox{0.95\columnwidth}{!}{%
  \includegraphics{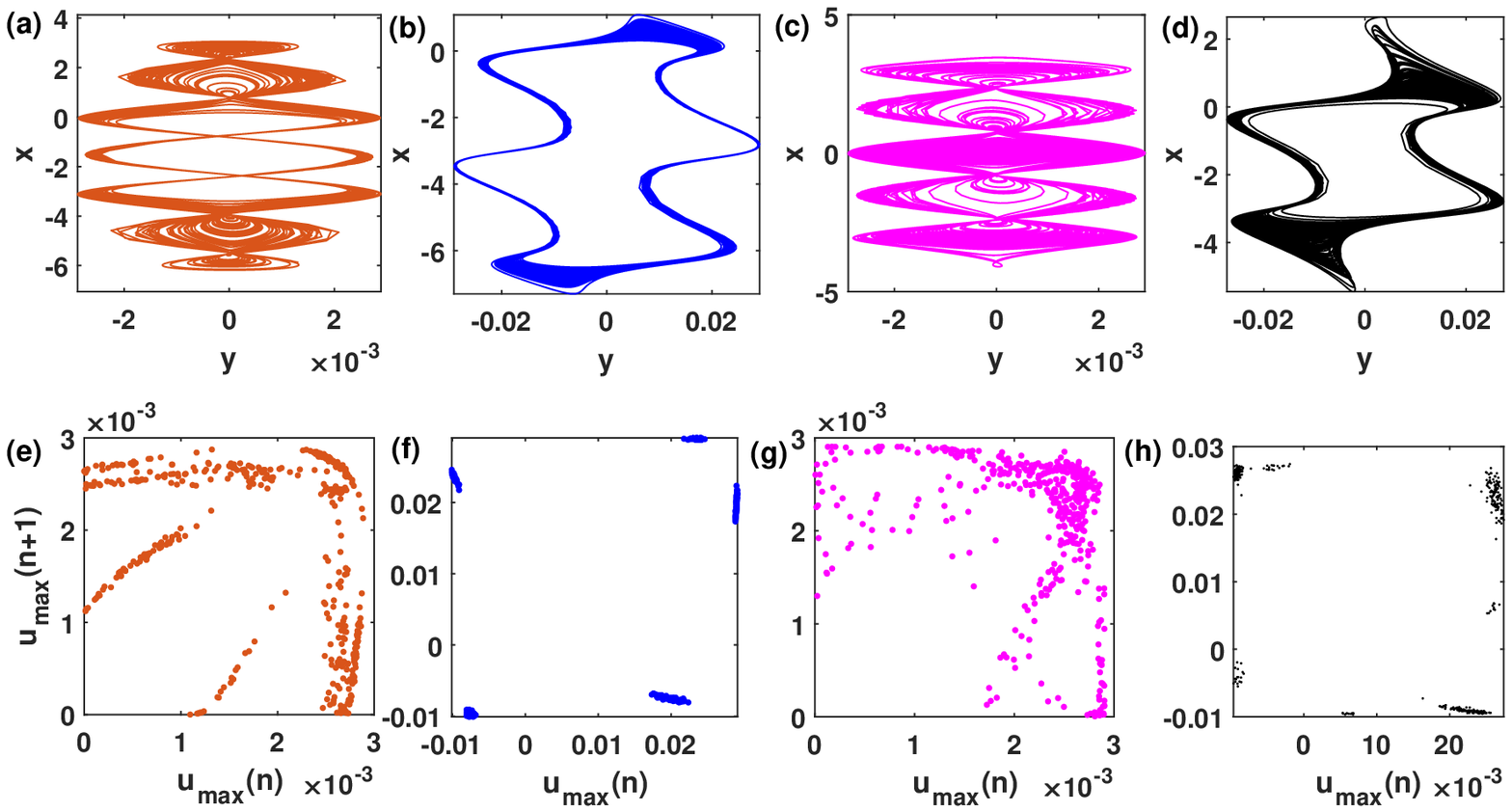} }
\caption{Phase space and max-map analysis: (a)-(d) respectively 2D phase portraits for the system (\ref{eq3}) with the respective $(K,a_2)=(0.01,1),~(0.01,1.5),~(0.06,1),~(0.045,1.5)$. The corresponding max maps are given in (e)-(f) respectively.}
\label{fig:3}       
\end{figure}

It can be observed from Fig.\ref{fig:2}(a) that, the values of $K_c \approx 1$ over $K \in [0.05,1]$ with $a_2=1.5$. Otherwise, $K_c \not \approx 1$ in the same $K_c$ vs. $K$ graph (see Fig.\ref{fig:2}(a)). It assures that chaotic dynamics exist in (\ref{eq3}) with $K \in [0.05,1]$, $a_2=1.5$. In Fig.\ref{fig:2}(b), it can be observed that values of $K_c \approx 1$ and $\not \approx 1$ with the respective $K \in [1,1.2]$ and $(1.2,1.5]$ for the fixed $K=0.045$. It confirms chaotic and non-chaotic dynamics of  sinusoidal blood flow in (\ref{eq3}) over $K \in [1,1.2]$ and $(1.2,1.5]$ respectively. We also investigate fluctuation in $K_c$ over the region $(K,a_2) \in [0,0.1] \times [1,1.5]$. The corresponding matrix plot is given in Fig.\ref{fig:2}(c). From the Fig.\ref{fig:2}(c), it can be seen that values of $K_c \approx 1$ for all most all $a_2 \in [1, 1.5]$ with $K>0.05$. It confirms chaos in (\ref{eq3}) over the parametric region: $[0.05,0.1] \times [1,1.5]$. However, different types of variation in $K_c (\not \approx 1)$ can be observed for the same $a_2$ with $K \leq 0.05$ (see the rectangular region marked by black dots). Fluctuation in $K_c$ is further investigate over the rectangular region given in Fig.\ref{fig:2}(d). From the figure, decreasing oscillation in $K_c$ can be observed for the increasing $a_2 \in [1,1.5]$ with $K \in [0,0.05]$. It confirms transition of chaotic to regular dynamics in (\ref{eq3}) with the increasing $a_2 \in [1,1.5]$ over $K \in [0.05,1]$. Further, it can be investigated from Fig.\ref{fig:2}(d) that the oscillation of $K_c$ always lies between $0.02$ to $0.08$ over $a_2 \in [1,1.5]$ with $K=0$. It verifies chaotic dynamics dynamics cannot be observed in the deterministic system (1). So, the whole study confirms that more complex as well as rich asymptotic dynamics can be observed in the stochastic red cell system compare to the same of its deterministic part.\vskip 3pt
We further investigate long-term dynamics in the system (\ref{eq3}) under the variation of $K \in [0,0.1]$ and $a_2 \in [1,1.5]$ respectively. Fig.\ref{fig:3}(a)-(d) shows some of the $2$D phase spaces with the respective $(K,a_2)=(0.01,1),~(0.01,1.5),~(0.06,1),~(0.045,1.5)$. 
From the figures, only periodic like movements in the trajectories can be observed in Fig.\ref{fig:3}(b) and (d) respectively. The remain shows intricate motion in the same (see Fig.\ref{fig:3}(a) and (c)). It indicates that the asymptotic dynamics in Fig.\ref{fig:3}(a),~(c), are more complicated than the same in Fig.\ref{fig:3}(b) and (d) respectively. Later, these are confirmed by clustering the patterns of the respective local-maxima maps. Local-maxima map (LMM) is defined as a Poincare$^{'}$ map of local maxima of a sequence $\{u(n)\}_{n=1}^{N}$ ($N$ being the length of time series.). It was established in \cite{REF7} that LMM can describe asymptotic dynamics of the system which even in the chaotic states. So, we investigate the nature of LMM for above phase spaces (shown in Fig.\ref{fig:3}(a)-(d) respectively). The corresponding LMMs are given in Fig.\ref{fig:3}(e)-(h) respectively. From the figures, it can be observed that the regular geometric structures exist in the LMM with the respective $K=0.01,~a_2=1.5$ and $K=0.045,~a_2=1.5$. It corresponds a periodic dynamics. On the other hand, remain shows no definite regular geometric structure exists in Fig.\ref{fig:3}(e) and (g). It confirms that the respective phase spaces are very complicated. The whole dynamical study is, thus assures the existence of chaos with complicates asymptotic dynamics in (\ref{eq3}). \par
In the next section, we quantify complexity of the phase spaces using heterogeneous recurrence based entropy (HRE)
\section{Complexity in the blood flow system and its quantification}
\label{sec:2}
\subsection{Methodology-Heterogeneous recurrence entropy (HRE)}
\label{sec:21}
Heterogeneous recurrence is generally defined on a phase space $P=(x_i) \in \mathbb{R}^d$ by measuring its local recurrences. The local recurrence is investigated by decomposing the phase space $P$ into some sub-regions. It can cluster the phase space and identify the heterogeneous recurrence patterns \cite{REF30,REF31,REF32,REF33,REF34}. In this context, $Q$-tree indexing method is one of the efficient tools that can be applied on large-scale multi-dimensional spatial data \cite{REF30,REF31,REF32,REF33,REF34}, and has a strong correlation with fractal structure \cite{REF30,REF31,REF32,REF33,REF34}. Further, $Q$-tree indexing divides $P$ into $2^d$ dimensional self-similar sub-regions constrain to a hierarchical structure in which the partition will continue until the number(states in a sub-regions) $\geq$ the capacity of a single state. So, each sub-regions can be treated as a time series which corresponds to a particular type of categorical variables. Such categorical variables can describe local evolution of the trajectories. Let $\vec{s}(n)$ denotes the sub-regions and $\Gamma=\{1,2,..,8\}$ be an index set. Then, $\vec{s}(n) \rightarrow k \in \Gamma$. by utilizing fractal theory, an IFS \cite{REF30,REF31,REF32,REF33,REF34} is then defined by
\begin{equation}\label{eq7}
\left[ \begin{array}{c} c_x(n) \\ c_y(n) \end{array} \right] =\phi \left(k,\left[\begin{array}{c} c_x(n-1) \\ c_y(n-1) \end{array} \right]\right)\ \begin{bmatrix} \alpha & 0 \\ 0 & \alpha \end{bmatrix} \times \left[ \begin{array}{c} c_x(n-1) \\ c_y(n-1) \end{array} \right]+ \left[\begin{array}{c} \cos(k \times \frac{2 \pi}{K}) \\ \sin(k \times \frac{2 \pi}{K}) \end{array} \right],
\end{equation}
where $\left[\begin{array}{c} c_x(0) \\ c_y(0) \end{array} \right]= \left[\begin{array}{c} 0 \\ 0 \end{array} \right]$ and $\alpha$ is a control parameter that prevents overlaps of two subregions in the graph \cite{REF30,REF31,REF32,REF33,REF34}. Here the map $\phi$ is a contractive mapping describes dynamics of $k \in \Gamma$ \cite{REF30,REF31,REF32,REF33,REF34}. Under the circular transformation (given by (\ref{eq7})), a fractal structure can be obtained, which contains $2^d$ recurrent structures. We denote $HR_{n_1}$, $HR_{n_1,n_2}$ for recurrent structure of individual state-$n_1$ and $2$-state $\vec{s}(n) \rightarrow n_1, \vec{s}(n-1) \rightarrow n_2$ respectively. Similarly, recurrent set of $M$-state $\vec{s}(n) \rightarrow n_1, \vec{s}(n-1) \rightarrow n_2,...,\vec{s}(n-M+1) \rightarrow n_M$ is denoted by $HR_{n_1,n_2,..,n_M}$. For any two elements $\psi^j,\psi^k \in HR_{n_1,n_2,..,n_M}$, a distance matrix $\mathcal{D}_{n_1,n_2,..,n_M}$ is then defined by $\mathcal{D}_{n_1,n_2,..,n_M}(j,k)=\|\psi^j- \psi^k\|$, where $j,k=1,2,..,L$ $(j < k)$. So, changes in $HR$ can be measured by variation in $\mathcal{D}$. The corresponding matrix plot of $\mathcal{D}$ is known as heterogeneous recurrence plot (HRP). \par
Utilizing Shannon entropy method, HRE is then defined by
\begin{equation}
H_{HR}=-\sum_{b=1}^{B}p(b) \log p(b),
\end{equation}
where $p(b)$ is given by
\begin{equation}
p(b)=\frac{1}{L(L-1)} \left(\frac{b-1}{B}~max(\mathcal{D}) <\mathcal{D}_{n_1,n_2,..,n_M} \leq \frac{b}{B}~max(\mathcal{D}) \right).
\end{equation}
Here, $b=1,2,..,B$ ($0 <B < max(\mathcal{D})$ being the number of bins).\par
In the following section, we investigate complexity in (\ref{eq3}) under the variation of $K \in [0,0.1]$ and $a_2 \in [1, 1.5]$.
\subsection{Quantifying complexity in (\ref{eq3}) by HRE }
\label{sec:21}
\begin{figure}[h!]
\resizebox{1.0\columnwidth}{!}{%
  \includegraphics{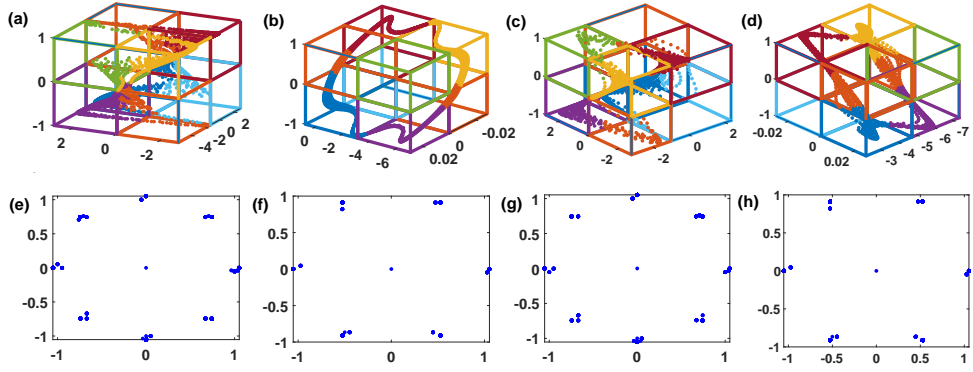} }
\caption{$Q$-index and IFS analysis: (a)-(d) respectively $2^3$-sub-devision for the phase spaces of (\ref{eq3}) with $(K,a_2)=(0.01,1),~(0.01,1.5),~(0.06,1),~(0.045,1.5)$. In each divisions are highlighted by different colors. The corresponding fractal IFS graphs are given in (e)-(f) respectively.}
\label{fig:4}       
\end{figure}

In this section, we first investigated HRP for for the sinusoidal blood flow in system  (\ref{eq3}) with $K \in [0,0.1]$ and $a_2 \in [1, 1.5]$. To construct HRP, we we make a partition of the phase space $P=\{(x,y,z) \in \mathbb{R}^3 \}$. Fig.\ref{fig:4}(a)-(d)  shows $2^3$ (as $d=3$) different sub-regions bounded by slef-similar cubes obtained by applying $Q$-indexing method with the respective $(K,a_2)=(0.01,1),~(0.01,1.5),~(0.06,1),~(0.045,1.5)$. We denote the sub-regions by $\vec{s}(n)$. Then by using $\vec{s}(n) \rightarrow k \in \Gamma$ and (\ref{eq7}), we construct categorical time series. The corresponding IFS fractal representation are given in Fig.\ref{fig:4}(e)-(h) respectively. From the Fig.\ref{fig:4}(e)-(h), it can be verified from the figures that each categories are lies within a unit circle centered around 8 addresses. In each cluster, the distribution of the recurrent points are varying individually. After the decomposition method stop, it is observed from the figures that Fig.\ref{fig:4}(e), (g) contain recurrent points that are much heterogeneous in nature compare to the same in Fig.\ref{fig:4}(f) and (h). It indicates more disordered asymptotic dynamics can be found in the case of $(K,a_2)=(0.01,1),~(0.06,1)$ compare to the same of $(K,a_2)=(0.01,1.5),~(0.045,1.5)$. 

\begin{figure}[h!]
\resizebox{1.0\columnwidth}{!}{%
  \includegraphics{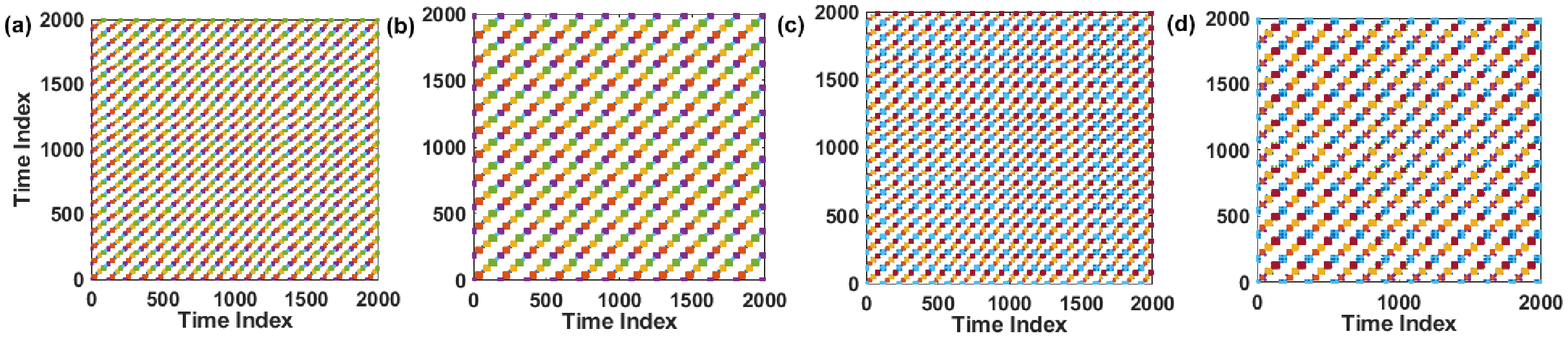} }
\caption{HRP analysis: (a)-(d) represents HRPs for the system (\ref{eq3}) with the respective $(K,a_2)=(0.01,1),~(0.01,1.5),~(0.06,1),~(0.045,1.5)$.}
\label{fig:5}       
\end{figure}

To investigate disorderness in the phase space, HRPs are constructed by the corresponding $\mathcal{D}_{n_1,n_2,..,n_M}$. For $(K,a_2)=(0.01,1),~(0.01,1.5),~(0.06,1),~(0.045,1.5)$, the respective HRPs are shown in Fig.\ref{fig:5}. 
From the Fig.\ref{fig:5}(a) and (c), different kind of heterogeneous recurrences can be observed. It indicates, existence of local transition between different states. On the other hand, very few heterogeneous structure can be observed in Fig.\ref{fig:5}(b) and (d). It implies, there exist very few local transitions between the states. As variation in $\mathcal{D}$ corresponds variation between the local transitions, it follows that disorder in the corresponding phase spaces shown in Fig.\ref{fig:3}(a), (d) are higher than the same given in Fig.\ref{fig:3}(b) and (d). In other words, more complex asymptotic dynamics can be observed in (\ref{eq3}) for the cases $(K,a_2)=(0.01,1),~(0.06,1)$ compare to the same for $(K,a_2)=(0.01,1.5),~(0.045,1.5)$.\par
\begin{figure}[h!]
\resizebox{1.0\columnwidth}{!}{%
  \includegraphics{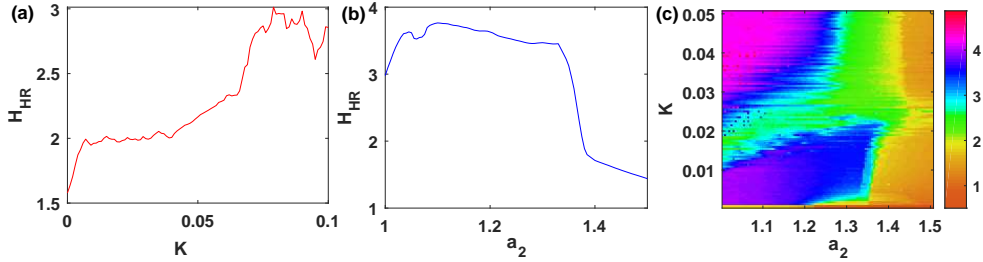} }
\caption{HRE analysis: (a) and (b) represents $H_{HR}$ vs. $K \in [0,0.1]$ ($a_2=1.5$) and $a_2 \in [1,1.5]$ ($K=0.045$) for the system (\ref{eq3})respectively. (c) represents $[H_{HR}(K,a_2)]$-matrix plot for the same system with $(K,a_2) \in [0,0.05] \times [1,1.5]$.}
\label{fig:6}       
\end{figure}

To quantify the complexity, we compute $H_{HR}$ with $K \in [0,0.1]$ ($a_2=1.5$) and $a_2 \in [1,1.5]$ ($K=0.045$). Fig.\ref{fig:6}(a) and (b) show respective fluctuations. 

From Fig.\ref{fig:6}(a), increasing $H_{HR}$ can be observed with the increase of $K \in [0,0.1]$ for fixed $a_2=1.5$. It confirms complexity in (\ref{eq3}) increases with the increasing $0<K \leq 0.1$ ($a_2=1.5$). On the other hand, both increasing and decreasing trend in the $H_{HR}$ can be observed for $a_2 \in [1,1.2]]$ and $(1.2,1.5]$ (fixed $K=0.045$) respectively. 
Further fluctuations in $H_{HR}$ over $(K,a_2) \in [0,0.05] \times [1,1.5]$ are computed. The corresponding matrix plot is given in Fig.\ref{fig:6}(c). From Fig.\ref{fig:6}(c), high and complex region can be investigated by measuring the values of $H_{HR}$. The blue, violet and red colors indicate $H_{HR}  \in [3,4.85]$. Further, it can be seen that sky blue, green, yellow indicate $H_{HR} \in [1,3)$. In this way, we have found complexity in (\ref{eq3}) over $(K,a_2) \in [0,0.05] \times [1,1.5]$. From the Fig.\ref{fig:6}(c), it can be also observed that the values of $H_{HR}$ lie between $0.5$ to $1.8$ for $a_2 \in [1,1.5]$ with fixed $K=0$. From remaining portion of the contour (shown in Fig.\ref{fig:6}(c)), it is observed that the values of $H_{HR}>1.5$ for some region of $a_2 \in [1,1.5]$ with the increasing $K>0$. It implies existence of greater complexity of the stochastic RBC system compare to the same of its deterministic part.
\section{Conclusion}
\label{sec:3}
In this study, chaotic dynamics of sinusoidal blood flow in the system (\ref{eq3}) has been found using $0-1$ test analysis. The corresponding numerical results have confirmed that both chaotic and non-chaotic states of sinusoidal blood flow exist in (\ref{eq3}) with the variation of noise strength ($K \in [0,0.1]$) and small axis ($a_2 \in [1,1.5]$). Further, the geometry of the system's long-term dynamics has been investigated by phase space and max-maps analysis. Both analyses have confirmed existence of regular and chaotic trajectories sinusoidal blood flow, respectively. It has been also observed that almost similar kind of transaction in the trajectories have been found in periodic as well as the chaotic states for sinusoidal blood flow. Using $H_{HR}$, we have quantified complexity of the phase spaces with the same variation in the respective $K$ and $a_2$. The quantification has successfully classified the stochastic blood flow system with a precise degree. In addition, more complex and rich dynamics have been quantified in the stochastic RBC system compare with its deterministic part. It indicates decreasing determinism in (\ref{eq1}) under the effect of noise and the small axis of the RBC.


\begin{thebibliography}{}
\bibitem{REF1} N. Mohandas, E. Evans, Annu. Rev. Biophys. Biomol. Struct. \textbf{23}, (1994) 787
\bibitem{REF2} R.Tran-Son-Tay, S. P. Sutera, P.R.Rao, Biophysical Journal \textbf{46}, (1984) 65-72
\bibitem{REF3} M. Abkarian, M. Faivre, A. Viallat, Phys. Rev. Lett. \textbf{98}, (2007) 188302
\bibitem{REF4} J. Dupire, M. Abkarian, A. Viallat, Phys. Rev. Lett. \textbf{104}, (2010) 168101
\bibitem{REF5} F. Takens, Lecture Notes in Mathematics \textbf{898}, (1981) 366-381
\bibitem{REF6} D. T. Kaplan, L. Glass, \textit{Understanding Nonlinear Dynamics} (Springer, New York, 1995)
\bibitem{REF7} S. H. Strogatz, \textit{Nonlinear Dynamics and Chaos} (Addison-Wesley, 1994)
\bibitem{REF8} E. Ott, \textit{Chaos in Dynamical Systems} (Cambridge University Press, 1993)
\bibitem{REF9} L. Rondoni, M. R. K. Ariffin, R. Varatharajoo, S. Mukherjee, S. K. Palit, S. Banerjee, Optics Communications \textbf{387}, (2017), 257-266
\bibitem{REF10} S. He, S. Banerjee, Physica A 501, (2018), 408–417
\bibitem{REF11} B. Yan, S. Mukherjee, S. He, Eur. Phys. J. Spec. Top. \textbf{228}, (2019) 2769-2777
\bibitem{REF12} S. Banerjee, M.R.K. Ariffin, Opt. Las. Tech. \textbf{45}, (2013) 435-442
\bibitem{REF13} S. K. Palit, N.A.A. Fataf, M.R. Md Said, S. Mukherjee, S. Banerjee, Eur. Phys. J. Spec. Top. \textbf{226}, (2017) 2219-2234
\bibitem{REF14} T. S. Dang, S. K. Palit, S. Mukherjee, T. M. Hoang, S. Banerjee, Eur. Phys. J. Spec. Top. \textbf{225}, (2016) 159-170
\bibitem{REF15} T. M. Hoang, S. K. Palit, S. Mukherjee, S. Banerjee, Optik \textbf{127} (2016) 10930-10947
\bibitem{REF16} G. A. Gottwald, I. Melbourne, Proc. R. Soc. Lond. A: Math., Phys. Engg. Sci. \textbf{460}, (2004) 603 
\bibitem{REF17} G. A. Gottwald, I. Melbourne, Phys. Rev. E \textbf{77}, (2008) 028201
\bibitem{REF18} C. Skokos, G.A. Gottwald, J. Laskar, Lecture Notes in Physics \textbf{915}, (2016) 221-247
\bibitem{REF19} G. A. Gottwald, I. Melbourne, Phys. D \textbf{212}, (2005) 100-110
\bibitem{REF20} D. Daems, G. Nicolis, Phys. Rev. E \textbf{59}, (1999) 4000-4006
\bibitem{REF21} A. N. Kolmogorov, IRE Trans. Inf. Theory \textbf{2}, (1956) 102-108
\bibitem{REF22} S. Mukherjee, S. K. Palit, S. Banerjee, M. R. K. Ariffin, L. Rondoni, D.K. Bhattacharya, Physica A \textbf{439}, (2015) 93-102
\bibitem{REF23} S. Banerjee, S. K. Palit, S. Mukherjee, M. R. K. Ariffin, L. Rondoni, Chaos \textbf{26}, (2016) 033105
\bibitem{REF24} S. Mukherjee, S. Banerjee, L. Rondoni, Physica A \textbf{508}, (2018) 131-140
\bibitem{REF25} Ya. G. Sinai, Dokl. Russ. Acad. Sci. \textbf{124}, (1959) 768-771
\bibitem{REF26} S. He, C. Li, K. Sun, S. Jafari, Entropy \textbf{20},(2018) 556
\bibitem{REF27} B. Yan, S. K. Palit, S. Mukherjee, S. Banerjee, Physica A \textbf{535}, (2019) 122433 
\bibitem{REF28} J.-P. Eckmann, S. O. Kamphorst, D. Ruelle, Europhysics Letters \textbf{4}, (1987) 973-977
\bibitem{REF29} N. Marwan, M. Carmen, Romano, M. Thiel, J. Kurths, Phys. Rep. \textbf{438}, (2007) 237-329
\bibitem{REF30} H. Yang, Y. Chen,  Chaos \textbf{24}, (2014) 013138
\bibitem{REF31} Y. Chen, H. Yang, European Physical Journal B \textbf{89}, (2016) 1-11
\bibitem{REF32} R. Chen, F. Imani and H. Yang, IEEE Journal of Biomedical and Health Informatics \textbf{24}, (2020) 1619-1631
\bibitem{REF33} C. Cheng, C. Kan, and H. Yang, Comput. Biol. Med. \textbf{75}, (2016) 10-18
\bibitem{REF34} H. Yang, C.-B. Chen, S. Kumara, Chaos \textbf{30}, (2020) 013119
\bibitem{REF35} K. L.-Hansen, V. V. Nikouline, J. M. Palva, R. J. Ilmoniemi, J. Neurosci. \textbf{21}, (2001) 1370-1377
\bibitem{REF36} F. N. Hooge, 1/f noises. Physica B+C \textbf{83}, (1976) 14-23
\end{thebibliography}
\end{document}